\documentclass[aps,twocolumn,amsmath,amssymb,prl,showpacs]{revtex4}
\usepackage{graphicx}
\usepackage[dvips,colorlinks=true,linkcolor=blue]{hyperref}

\begin{document}
\title{Understanding Anomalous Transport in Intermittent Maps:\\
From Continuous Time Random Walks to Fractals}
\author{N. Korabel$^{1}$, A.V. Chechkin$^{2}$, R. Klages$^{1}$,
I.M. Sokolov$^{3}$, V.Yu. Gonchar$^{2}$} 
\affiliation{$^1$Max-Planck-Institut f\"ur Physik komplexer Systeme,
N\"othnitzer Str.\ 38, D-01187 Dresden, Germany\\ 
$^2$Institute for Theoretical Physics NSC KIPT, Akademicheskaya st.\ 1, 61108 Kharkov, Ukraine\\
$^3$Institut f\"ur Physik, Humboldt-Universit\"at zu Berlin, Newtonstr.\ 15,
D-12489 Berlin, Germany}  
\date{\today}
\begin{abstract}
We show that the generalized diffusion coefficient of a subdiffusive
intermittent map is a fractal function of control parameters. A modified
continuous time random walk theory yields its coarse functional form and
correctly describes a dynamical phase transition from normal to anomalous
diffusion marked by strong suppression of diffusion. Similarly, the
probability density of moving particles is governed by a time-fractional
diffusion equation on coarse scales while exhibiting a specific fine
structure. Approximations beyond stochastic theory are derived from a
generalized Taylor-Green-Kubo formula.
\end{abstract}

\maketitle

The notion of {\it anomalous diffusion} derives from the fact that the mean
squared displacement (MSD) must not follow the law of normal diffusion, $<x^2>
\sim t^\gamma$ with $\gamma = 1$, but may be superdiffusive with $\gamma > 1$
or subdiffusive with $\gamma < 1$. Such anomalous dynamics has been
investigated theoretically and observed experimentally not only in amorphous
semiconductors, surface diffusion, turbulence, polymers and plasmas but also
in chemical, biological and economical problems
\cite{Bou90,Met00}. For this variety of applications it is desirable to have
a class of test systems at hand which are yet easy to handle. Here
periodically continued deterministic maps provide an important basis for
analytical and numerical investigations
\cite{Gei84,Gei85,Zum93,Zum93a,Sto95,Bar03}. Both the sub-
and the superdiffusive dynamics of such models was successfully described by
continuous time random walk (CTRW) approaches
\cite{Gei84,Zum93,Zum93a,Bar03}. This stochastic theory yields
non-Gaussian probability density functions (PDF) for anomalous diffusive
processes exhibiting stretched exponential decay in case of sub- and L{\'e}vy
power law decay in case of superdiffusion \cite{Zum93}.

On the other hand, a {\em microscopic theory} of anomalous deterministic
transport explaining the origin of this behavior in terms of the theory of
dynamical systems is just beginning to evolve \cite{Sto95}. Surprises along
these lines were already encountered in normal diffusive maps, where the
diffusion coefficient was found to be a fractal function of control parameters
\cite{Kla95,Kla96}. This behavior was also reported for more complex systems
like the climbing sine map \cite{Kor02}, bouncing ball billiards \cite{Har01}
and coupled Josephson junctions \cite{Tan02}. The fractality can be understood
as a signature of long-range dynamical correlations that, due to topological
instabilities, change in a complicated manner under parameter variation
\cite{Kla02}. However, no such fractal structures were yet identified in
anomalous dynamics as modeled by the maps of Refs.\
\cite{Gei84,Gei85,Zum93,Zum93a,Sto95,Bar03}.

\begin{figure}[b]
\includegraphics[width=8.5cm]{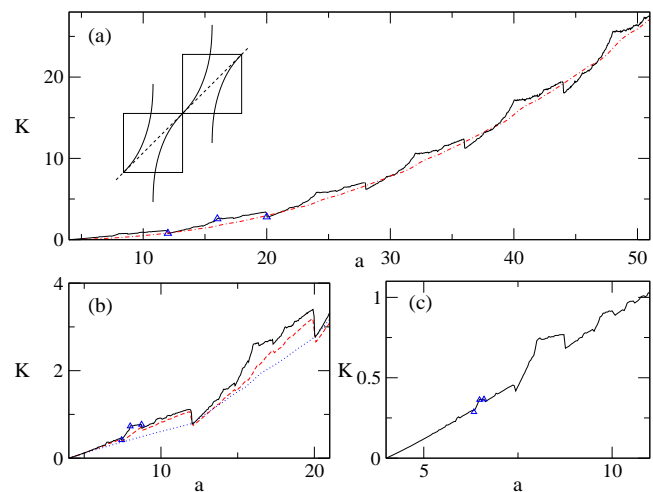}
\caption{The generalized diffusion coefficient $K$, Eq.\ (\ref{GDC_Def}),
as a function of $a$ for $z=3$. The curve in (a) consists of $1200$ points,
the dashed-dotted line displays the CTRW result $K_1$, Eqs.\
(\ref{jump_length}), (\ref{GDC_CTRW}). (b) (600 points) and (c) (200 points)
show magnifications of (a) close to the onset of diffusion. The dotted line in
(b) is the CTRW approximation $K_2$, Eqs.\ (\ref{int_jump_length}),
(\ref{GDC_CTRW}). The dashed line represents the first term of the TGK formula
Eq.\ (\ref{GDC_GK}). The triangles mark a specific structure appearing on
finer and finer scales. The inset depicts the model Eq.\ (\ref{map}). All
quantities here and in the following figures are dimensionless.}
\label{Fig1}
\end{figure}

In this Letter we focus on a subdiffusive map whose functional form on the
unit interval was introduced by Pomeau and Manneville for describing
intermittency \cite{Pom80},
\begin{equation}
\label{map}
x_{n+1}\equiv M_{a,z}(x_n) = x_n +a x_n^z,\;\; 0 \le x_n <\frac 12 .
\end{equation}
For $\frac 12\le x_n<1$ the map is defined according to
$M_{a,z}(-x)=-M_{a,z}(x)$. The translation $M_{a,z}(x + m) = M_{a,z} (x) + m$,
$m \in \mathbb{Z}$, completes the definition on the real line, see the inset
of Fig.\ \ref{Fig1} (a).  The degree of nonlinearity $z\ge 1$ and $a
\ge 1$ in Eq.\ (\ref{map}) hold for the two control parameters. For $1 \le z
< 2$ this model leads to normal diffusion while for $z \ge 2$ its behavior is
anomalous \cite{Gei84,Zum93}. The anomalous regime results from the existence
of marginal fixed points located at all integer values of $x$. Thus, a typical
trajectory of the map consists of long laminar phases interrupted by chaotic
bursts. In contrast to Refs.\ \cite{Gei84,Zum93}, which focused on the time
dependence of the MSD for the particular parameter value of $a=2^z$, here we
study the behavior of the generalized diffusion coefficient (GDC) \cite{Met00}
\begin{equation}  
\label{GDC_Def}
K := \lim_{n \rightarrow \infty} \frac{\left< x^2 \right>}{n^{\gamma}}\:
\end{equation}
($<\ldots>$ denotes an ensemble average) for this map under variation of both
control parameters.  We first focus on numerical simulations \cite{sim} of the
map Eq.\ (\ref{map}) and on the interpretation of the results within the CTRW
approach.  CTRW theory models diffusion processes by sequences of jumps
interrupted by periods of waiting. Let the PDFs of waiting times and of jump
lengths be defined by $\phi(t)$ and $\lambda(x)$, respectively. Choosing these
functions appropriately, CTRW theory predicts that $\gamma=1$ for $1\le z<2$
and $\gamma=1/(z-1)$ for $2<z$ irrespective of the parameter $a$ of the map
\cite{Gei84,Zum93}. Indeed, for all $a$ we find excellent agreement between
these solutions and the values for $\gamma$ obtained from
simulations. Consequently, the analytical expressions for $\gamma$ are used
throughout our work.

\begin{figure}[t]
\includegraphics[width=8.5cm]{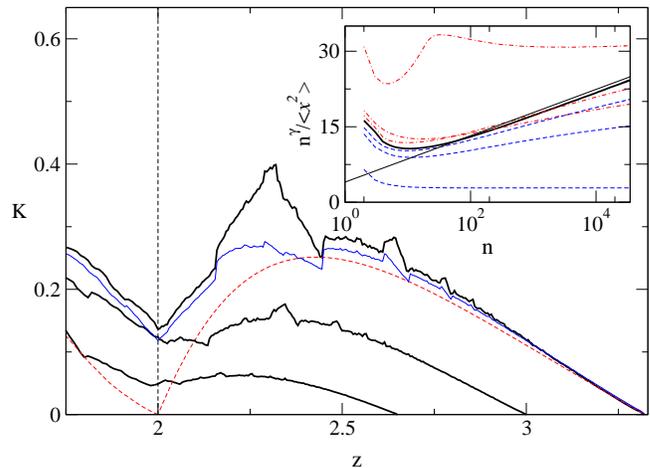}
\caption{$K$ as a function of $z$ for $a=3.14, 4, 5$ consisting of $200$,
$250$ and $300$ points (bottom to top). Bold lines represent simulation
results for $n=10^4$ iterations, the dashed line is the CTRW solution $K_2$,
Eqs.\ (\ref{int_jump_length}), (\ref{GDC_CTRW}), for $n\to\infty$. The thin
line displays the first term of the TGK formula Eq.\ (\ref{GDC_GK}). At $z=2$
there is a transition from normal to anomalous diffusion. The inset describes
the time-dependence of $<x^2>$ around this value with $a=3.14$. Here dashed
lines correspond to $z=1.5, 1.9, 1.95$ (bottom to top), dashed-dotted lines to
$2.5, 2.1, 2.05$ (top to bottom) while the bold line depicts $z=2$. The dotted
line is proportional to $\ln(n)$.}
\label{Fig2}
\end{figure}

Simulation results for $K$ as a function of $a$ for fixed $z$ are presented in
Fig.\ \ref{Fig1}. Magnifications of Fig.\ \ref{Fig1} (a) shown in parts (b)
and (c) reveal self similar-like irregularities indicating a fractal parameter
dependence of $K$. Note particularly the structure marked by triangles, which
is repeated on finer and finer scales. The parameter values for these symbols
correspond to specific series of Markov partitions \cite{Kla95,Kla96}. In
Fig.\ \ref{Fig2} $K$ is depicted as a function of $z$ for different values of
$a$, again displaying highly non-monotonic parameter dependences.

In a first step for explaining these curves we apply standard CTRW theory
\cite{Gei84,Zum93} by modifying this approach at
three points: Firstly, the waiting time PDF must be calculated according to
the grid of elementary cells indicated in Fig.\
\ref{Fig1} \cite{Kla96,Kla97} yielding
\begin{equation}
\phi (t)=a\left( 1+a(z-1)t\right) ^{-\frac z{z-1}}\:.
\label{waiting_time_pdf}
\end{equation}
Secondly, as a jump probability we use
\begin{equation}  
\label{real_jump_PDF}
\lambda (x) = \frac{p}{2} \delta (\left| x \right| - l) + (1 - p) \delta (x)\:.
\end{equation}
The second term accounts for the fact that particles may stay in a box with a
probability of $(1 - p)$. This quantity is roughly determined by the size of
the escape region $p = ( 1 - 2 x_c )$ with $x_c$ as a solution of the equation
$x_c + a x^{z}_{c} = 1$. Thirdly, we introduce two definitions of a typical
jump length $l_i\:,i\in\{1,2\}$. Thus,
\begin{equation}  
\label{jump_length}
l_1 = \left\{ | M_{a,z}(x) - x | \right\} \:
\end{equation}
corresponds to the actual mean displacement while

\begin{equation}  
\label{int_jump_length}
l_{2} = \left\{ | [M_{a,z}(x)] | \right\} \:
\end{equation}
gives the coarse-grained displacement in units of elementary cells, as it is
often assumed in CTRW approaches. In these definitions $\left\{\ldots\right\}$
denotes both a time and ensemble average over particles leaving a box. The
modified CTRW approximation for the GDC resulting from these three changes
reads
\begin{equation}
K_i = \begin{cases} 
p l_i^2 a^{\gamma} \sin (\pi \gamma) / \pi \gamma^{1+\gamma}, & 0 < \gamma < 1
\cr  
p l_i^2 a (1-1/\gamma), & 1 \le \gamma < \infty \: .
\end{cases}
\label{GDC_CTRW}
\end{equation}
Fig.\ \ref{Fig1} (a) shows that $K_1$ well describes the coarse functional
form of $K$ for large $a$. $K_2$ is depicted in Fig.\ \ref{Fig1} (b) by the
dotted line and is asymptotically exact in the limit of $a\to0$. Hence, the
GDC exhibits a dynamical crossover analogous to the one found for normal
diffusion \cite{Kla96,Kla97,Kor02}. Let us now focus on the GDC at
$a=12,20,28,...$, which corresponds to integer values of the height
$h=[M_{a,z}(1/2)]$ of the map. Simulations reproduce, within numerical
accuracy, the results for $K_2$ by indicating that $K$ is discontinuous at
these parameter values. Due to the self similar-like structure of the GDC we
arrive at the conjecture that $K$ exhibits infinitely many discontinuities on
fine scales as a function of $a$, in contrast to the continuity of the CTRW
solution.

We now switch back to the $z$-dependence of $K$, where CTRW theory predicts a
dynamical phase transition from normal to anomalous diffusion at $z=2$
\cite{Gei84,Zum93}.  At the transition point one has $<x^2>\sim
n/\log(n)$. According to Eq.\ (\ref{GDC_Def}), Eq.\ (\ref{GDC_CTRW}) yields a
continuous transition in $z$, see $K_2$ in Fig.\ \ref{Fig2}. However, this
coarse functional form is obscured not only by fractal irregularities but also
by an ultra-slow convergence of the MSD in our simulations. Using CTRW theory,
we verified that {\em around} $z=2$ the MSD is determined by a series of
logarithmic corrections in time, where the lowest-order term is proportional
to $n^{\gamma}/\log n$ and dominates at time scales controlled by
$n\ll\tilde{n}\sim \exp(1/(1-\gamma))$. Hence, for $\gamma\neq 1$ and
$n\gg\tilde{n}$ simple power laws are recovered while for $\gamma=1$ this
logarithmic term survives in the infinite time limit. The inset of Fig.\
\ref{Fig2} exemplifies this behavior explaining deviations between CTRW theory
and the simulation results in the main figure. In other words, around $z=2$
both normal and anomalous diffusion are suppressed due to logarithmic
corrections in time leading to a vanishing $K$. We strongly suspect that such
a behavior of the GDC is typical for dynamical phase transitions in anomalous
dynamics altogether \cite{Zum93a}.

The stochastic CTRW theory presented above describes only the coarse parameter 
dependence of the GDC. This motivates us to develop an approach that incorporates
dynamical correlations. For this purpose we generalize the
Taylor-Green-Kubo formula (TGK) \cite{Remark2} for maps to anomalous
diffusion. We start by expressing Eq.\ (\ref{GDC_Def}) via sums over the
integer velocities $v_m=\left[x_{m+1}\right]-\left[x_m\right]$
\cite{Kla02}. However, in contrast to normal diffusion the anomalous dynamics
generated by Eq.\ (\ref{map}) is not stationary for $z \ge 2$
\cite{Gas88,Zum93a}. This is reminiscent in the non-existence of an invariant
probability density for this map, or mathematically, in the existence of
infinite invariant measures \cite{Aar}. Consequently, the resulting expression
cannot be simplified using time-translational invariance, and we get
\begin{equation}  
K = \lim_{n
\rightarrow \infty} \frac{1}{n^{\gamma}} \left[ <\sum_{k=0}^{n-1} v_k^2> + 2
<\sum_{k=0}^{n-1} \sum_{l=1}^{n-1} v_k v_{k+l}> \right] \:.
\label{GDC_GK}
\end{equation}
Focusing on the first term only, simulations confirm that it is proportional
to $n^{\gamma}$. For $z<2$ the dynamics is ergodic, hence this term is equal
to $n <v_0^2>$, and we recover the random walk formula for normal diffusion
\cite{Kla96,Kla97}. For $z\ge 2$ only generalized ergodic theorems may hold
\cite{Aar}, and whether the CTRW result Eq.\ (\ref{GDC_CTRW}) can be derived
from the first term is a non-trivial question. Considering this term as an
approximation of $K$ for all $z$, the numerical results are depicted in Figs.\
\ref{Fig1} (b) and \ref{Fig2}. The comparison of this data with the simulation
values based on Eq.\ (\ref{GDC_Def}) indeed shows that this term already
provides a first step beyond the modified CTRW model: In the $a$ dependence it
reproduces the major irregularities of $K$ and even appears to follow the
discontinuities discussed above. However, we cannot say yet whether it yields
exact values for $K$ at integer heights.

Eq.\ (\ref{GDC_GK}) thus provides a suitable starting point for a systematic
understanding of the fractal GDC beyond CTRW theory. We remark that already
the first term can be related to de Rham-type fractal functions, which
explains why it features irregularities \cite{Unp}. Since the series expansion
in Eq.\ (\ref{GDC_GK}) is exact, working out further terms one must recover
more and more structure in the GDC \cite{Kla96,Kla02,Kor02}. Note that the
correlation function in the second term depends on two times, which enables to
express $K$ in terms of {\em aged} \cite{Bar03} de Rham-type fractal functions
summing up to the exact value.
\begin{figure}[t]
\includegraphics[width=8cm]{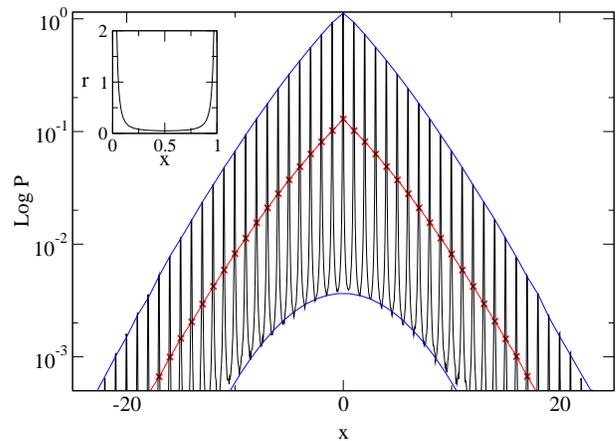}
\caption{Comparison of the PDF obtained from simulations of the map Eq.\
(\ref{map}) (oscillatory structure) with the analytical solution Eq.\
(\ref{Sol}) of the fractional diffusion equation Eq.\ (\ref{Frac_Dif_Eq})
(continuous line in the middle) for $z=3$ and $a=8$. The PDF was computed from
$10^7$ particles after $n=10^3$ iterations. For the GDC in Eq.\ (\ref{Sol})
the simulation result was used. The crosses (x) represent the numerical
results coarse grained over unit intervals. The upper and the lower curves
correspond to fits with a stretched exponential and a Gaussian distribution,
respectively. The inset depicts the PDF for the map on the unit interval with
periodic boundaries.}
\label{Fig3}
\end{figure}

We now turn to the PDFs generated by the map Eq.\ (\ref{map}). The great
success of CTRW theory derives from the fact that it correctly predicts both
the power $\gamma$ and the form of the coarse grained PDF $P(x,t)$ of
displacements for a large class of models \cite{Zum93}. Correspondingly, the
diffusion process generated by Eq.\ (\ref{map}) is not described by an
ordinary diffusion equation but by a fractional generalization of it. Starting
from the CTRW model for the map Eq.\ (\ref{map}) discussed above, one can
derive the time-fractional diffusion equation
\begin{equation}
\frac{\partial ^\gamma P(x,t)}{\partial t^\gamma }\equiv
\int_0^tdt^{^{\prime }}(t-t^{^{\prime }})^{-\gamma }\frac{\partial P}{%
\partial t^{^{\prime }}}=D\frac{\partial ^2P}{\partial x^2} 
\label{Frac_Dif_Eq}
\end{equation}
describing the long-time limit of the PDF of the coarse grained dynamics with
initial condition $P(x,0)=\delta(x)$ and $D=K\Gamma (1 +
\gamma)/2$, $0<\gamma <1$. The fractional derivative is understood in
the Caputo sense \cite{Mai97}. Time-fractional equations of such a form have
already been extensively studied by mathematicians \cite{Gor00}. We remark
that the two other fractional diffusion equations proposed in Refs.\
\cite{Bar03,Met00}, which are based on a Riemann-Liouville fractional
derivative, are equivalent to Eq.\ (\ref{Frac_Dif_Eq}) under rather weak
assumptions \cite{ACunp}. The solution of Eq.\ (\ref{Frac_Dif_Eq}) expressed
in terms of an M-function of Wright type
\cite{Mai97,Gor00} reads
\begin{equation}
P(x,t)=\frac 1{2\sqrt{D}t^{\gamma /2}}M\left( \xi,\frac \gamma 2\right)
\label{Sol}
\end{equation}
giving exactly the same asymptotics that was obtained in Ref.\ \cite{Zum93}
for small and large values of $\xi =|x|/\sqrt{D}t^{\gamma/2}$. By using a
series representation of $M$ it can be demonstrated \cite{ACunp} that this
form is equivalent to those expressed via H-functions \cite{Met00} or
one-sided extremal L\'evy stable distributions \cite{Bar03}. Fig.\ \ref{Fig3}
demonstrates an excellent agreement between the analytical solution Eq.\
(\ref{Sol}) and the PDF obtained from simulations for the map Eq.\ (\ref{map})
if the PDF is coarse grained over unit intervals. However, it also shows that
the coarse graining eliminates a periodic fine structure that is not captured
by Eq.\ (\ref{Sol}) \cite{Kla96}. This fine structure derives from the
`microscopic' PDF of an elementary cell (with periodic boundaries) as
represented in the inset of Fig.\ \ref{Fig3}. The singularities are due to the
marginal fixed points of the map, where particles are trapped for long
times. Remarkably, that way the microscopic origin of the intermittent
dynamics is reflected in the shape of the PDF on the whole real line: From
Fig.\ \ref{Fig3} it is seen that the oscillations in the PDF are bounded by
two functions, the upper curve being of a stretched exponential type while the
lower is Gaussian. These two envelopes correspond to the laminar and chaotic
parts of the motion, respectively \cite{Remark3}.

In conclusion, we have shown that the anomalous dynamics generated by a
paradigmatic subdiffusive one-dimensional map exhibits a fractal GDC under
variation of control parameters. The coarse dependence of this GDC and a
non-trivial phase transition from normal to anomalous diffusion are captured
by a modified CTRW theory. Near the phase transition point the GDC is strongly
suppressed by logarithmic corrections in time. A more detailed understanding
of the GDC is provided by an anomalous TGK formula suggesting intimate
relations to aging, fractal functions and ergodic theory. The coarse-grained
PDF of this anomalous dynamics is in excellent agreement with the solution of
a suitable fractional diffusion equation while on fine scales it reflects the
microscopic details of the intermittent dynamics. Here we have only treated a
subdiffusive map, however, we expect these findings to be typical for
spatially extended, low-dimensional, anomalous deterministic dynamical systems
altogether. Further studies will focus on a more detailed analysis of the
anomalous TGK-formula, on a spectral analysis of this map and on treating
superdiffusive dynamics along the same lines.

The authors acknowledge helpful discussions with J. Klafter, R. Metzler and
A. Pikovsky. They thank the MPIPKS Dresden for hospitality and financial
support. IMS acknowledges partial financial support by the Fonds der
Chemischen Industrie.


\begin{thebibliography}{Bou90}
\bibitem{sim} All simulations were performed starting from a uniform,
random distribution of $10^6$ initial conditions on the unit interval by
iterating for $n=10^4$ time steps.
\end{thebibliography}
\end{document}